\documentclass[letterpaper, 10 pt, conference]{ieeeconf}  

\IEEEoverridecommandlockouts                              
\usepackage{xcolor}
\usepackage{float}
\usepackage{amsmath}

\usepackage{amsthm,amssymb,graphicx,url}
\let\labelindent\relax 
\usepackage{enumitem}
\usepackage{booktabs}
\usepackage{caption}
\usepackage{subcaption}

\usepackage{appendix}

\usepackage{wrapfig}
\usepackage{hyperref}
\usepackage{soul}  
\usepackage{cleveref}
\usepackage{bm}
\usepackage{multicol}
\usepackage{mathtools}
\usepackage{cite}
\usepackage{graphicx}

\crefname{equation}{}{} 
\crefname{assumption}{Assumption}{}
\crefname{table}{Table}{} 
\crefname{figure}{Fig.}{}
\crefname{section}{Section}{}
\crefname{remark}{Remark}{}
\usepackage{algorithm,algorithmicx} 
\usepackage{algpseudocode} 
\newlength\myindent
\DeclareMathOperator*{\argmin}{argmin}

\def\trieq{\triangleq}

\newtheorem{theorem}{Theorem}
\newtheorem{lemma}{Lemma}
\theoremstyle{definition}  
\theoremstyle{definition} \newtheorem{assumption}{Assumption}
\theoremstyle{remark}  \newtheorem{remark}{Remark}

\title{\LARGE \bf
Time Coordination of Multiple UAVs over Switching Communication Networks with Digraph Topologies
}

\author{Hyungsoo Kang, Hyung-Jin Yoon, Venanzio Cichella,  Naira Hovakimyan, and Petros Voulgaris
\thanks{This work is supported by AFOSR.}
\thanks{Hyungsoo Kang and Naira Hovakimyan are with the Department of Mechanical Science and Engineering, University of Illinois at Urbana-Champaign, 
        Urbana, IL 61801, USA.  
        {\tt\small \{hk15, nhovakim\} @illinois.edu}}

\thanks{Hyung-Jin Yoon and Petros Voulgaris are with the Department of
Mechanical Engineering, University of Nevada, 
        Reno, NV 89557, USA.
        {\tt\small \{hyungjiny, pvoulgaris\}@unr.edu}}
        
\thanks{Venanzio Cichella is with the Department of Mechanical
Engineering, University of Iowa, 
        Iowa City, IA 52242, USA.
        {\tt\small venanzio-cichella@uiowa.edu}}        
}

\begin{document}

\maketitle
\thispagestyle{empty}
\pagestyle{empty}

\begin{abstract}
This paper presents a time-coordination algorithm for multiple UAVs executing cooperative missions. Unlike previous algorithms, it does not rely on the assumption that the communication between UAVs is bidirectional. Thus, the topology of the inter-UAV information flow can be characterized by digraphs. To achieve coordination with weak connectivity, we design a switching law that orchestrates switching between jointly connected digraph topologies. In accordance with the law, the UAVs with a transmitter switch the topology of their coordination information flow. A Lyapunov analysis shows that a decentralized coordination controller steers coordination errors to a neighborhood of zero. Simulation results illustrate that the algorithm attains coordination objectives with significantly reduced inter-UAV communication compared to previous work.
\end{abstract}

\section{INTRODUCTION}
In recent years, the field of multi-vehicle control has undergone extensive research and development in order to address a variety of challenging problems. Relevant examples include cooperative payload transportation with UAVs \cite{hye2017,tae2018}; cooperative simultaneous localization and mapping (SLAM), where a team of UAVs cooperatively constructs or updates a map of an unknown large area \cite{isa2016,pat2017}; rescue and surveillance missions \cite{tri2018,sat2017}.

Of the diverse topics in cooperative multi-UAV systems, 
time coordination of multiple UAVs has been an area of increasing importance because it determines the safety and efficacy of a cooperative mission. Its representative applications are sequential auto-landing and  simultaneous suppression of multiple ground targets. At the planning stage of a cooperative mission, a trajectory generation algorithm \cite{cho2016,aug2012} designs a set of desired collision-free trajectories together with a set of desired speed profiles. A path-following controller \cite{lap2006,gha2007,ven2011} allows each UAV to follow its virtual target which defines the desired position of it and slides along the trajectory in accordance with the speed profile. However, when the mission unfolds, disturbances such as wind gusts and temporary hardware failure may put some UAVs behind or ahead of their virtual targets, thereby causing inter-UAV discoordination and jeopardizing the success of the mission. To restore coordination, a coordination algorithm~adjusts~the progression of the virtual targets using coordination information exchanged between the UAVs over a time-varying bidirectional network \cite{xar2013,ven2015}. The research in \cite{pui2015} additionally studies absolute temporal requirements, e.g., arrival of the UAVs within a prescribed time range. In \cite{cam2020,cam2021}, the authors present time-coordination algorithms which can achieve obstacle avoidance as well. However, the time-coordination algorithms in these studies restrict the topologies of the dynamic communication network to bidirectional graphs, a special class of directed graphs. In other words, the algorithms require the communication between two UAVs to be bidirectional. This is because the stability analysis of these algorithms relies on the symmetricity of the Laplacian. Thus, they cannot be applied to dynamic directional inter-UAV communication cases, where the symmetricity of the Laplacian is no longer guaranteed. A different approach has to be taken to address the time-coordination problem over a dynamic directional communication network, which motivates our present work.

Recent studies show that strategic switching of the network topology plays an important role in solving control problems of networked multi-agent systems. For example, \cite{gua2005} presents a centralized topology switching algorithm to achieve consensus for the first-order multi-agent systems. Active topology switching algorithms proposed in \cite{yan2018a,yan2018b} achieve consensus for the second-order multi-agent systems. In \cite{ert2017}, a dynamic network topology control
problem in adversarial environments is presented, where the network designer strategically changes the topology to yield desirable network properties, while an adversary tries to damage the network functionality. Also, considerable advances in wireless communication technology have made it more feasible to set the topology of the inter-UAV communication network as a control variable \cite{maz2011}.

The contributions of this paper can be summarized as follows. 1) Inspired by  state-feedback switched system theory in \cite{zhe2005}, we design a switching law for the directional inter-UAV communication network, over which a decentralized coordination controller solves the time-coordination problem. The law does not require the network to be connected via~a directed spanning tree at any time instant.
With Jointly connected communication, a small number of communication
edges are activated at each time instant consuming a short
portion of the limited bandwidth.
2) Since the information flow between UAVs is directional, not all the UAVs need to be equipped with both a transmitter and a receiver. For the same reason, the amount of inter-UAV communication required to solve the problem can be significantly reduced compared to the bidirectional case~\cite{ven2015}.
Thus, one can cut costs on communication devices and save consumption of the bandwidth and energy for communication.

The rest of the paper is organized as follows. In Section~\ref{prelim}, basic definitions and an algebraic digraph theory~are given. Section \ref{III} introduces the time-coordinated path-following framework that lays the basis for the problem formulation. Section \ref{IV} provides key assumptions on the inter-UAV communication and describes the time-coordination problem. In section \ref{V}, we propose a decentralized coordination controller and design a switching law for the communication network, followed by a presentation of the main results of this paper. Section \ref{VI} reports simulation results. Finally, Section \ref{VII} summarizes the paper.
\section{PRELIMINARIES} \label{prelim}
\subsection{Graph Theory}
A digraph of order $n$ is defined as $\mathcal{D}=(\mathcal{V},\mathcal{E},\mathcal{A})$, where $\mathcal{V}=\{1,\dots,n\}$ is the set of nodes, $\mathcal{E}\subset \mathcal{V}\times \mathcal{V}$ is the set of edges of $\mathcal{D}$, and $\mathcal{A}$ is the adjacency matrix of $\mathcal{D}$. A directed edge $(i,j)\in\mathcal{E}$ means that information can be transmitted from node $j$ to node $i$. The adjacency matrix $\mathcal{A}$ is defined as $[\mathcal{A}]_{ij}=1$, if $(i,j)\in\mathcal{E}$ and $[\mathcal{A}]_{ij}=0$, otherwise. The neighborhood of node $i$ is the set $\mathcal{N}_i=\{j\in\mathcal{V} : \ (i,j)\in\mathcal{E}\}$. The Laplacian of $\mathcal{D}$ is $L=\Delta-\mathcal{A}$, where $\Delta=diag\{d_1,d_2,\dots,d_n\}$ and $d_i=\sum_{j\in\mathcal{N}_i}[\mathcal{A}]_{ij}$ is the in-degree of node $i$. Based on the structure of $L$, at least one of its eigenvalues is located at $0$ and the rest of them lie in the right half plane. A directed path from node $i_s$ to $i_0$ is a sequence of directed edges $(i_0,i_1)$, $(i_1,i_2)$, $\dots$,~$(i_{s-1},i_s)$. If there exists a node such that every other node is reachable along a directed path from it, the digraph is said to contain a directed spanning tree or to be connected via a directed spanning tree. If a digraph $\mathcal{D}$ contains a directed spanning tree, $L$ has a simple eigenvalue $0$ with the corresponding eigenvector $1_n$. 
Otherwise, the multiplicity of the eigenvalue $0$ of $L$ is greater than one.


\section{TIME-COORDINATED PATH-FOLLOWING FRAMEWORK} \label{III}
\begin{figure} [t!]
    \centering
    \includegraphics[width = 1.00\linewidth]{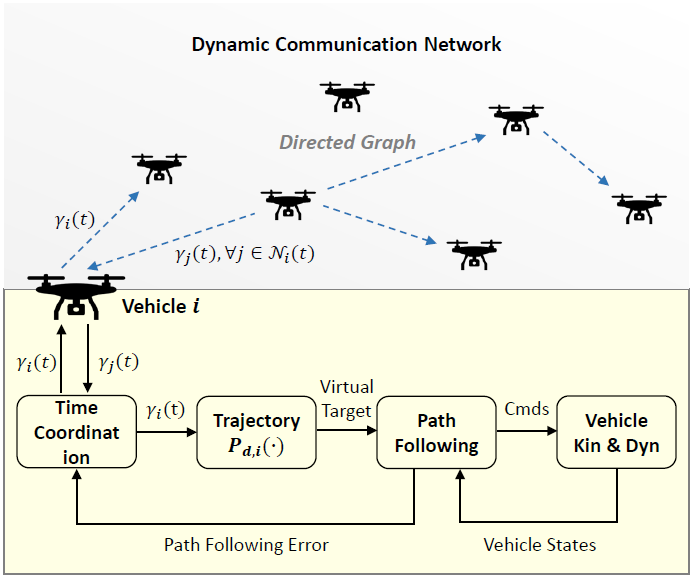}
    \caption{Time-coordinated path following of multiple UAVs. The time-coordination module governs the evolution of the virtual time $\gamma_i(t)$, thereby adjusting the progression of the UAVs along the desired trajectories to achieve intervehicle coordination in combination with a path-following controller.}
    \label{fig:architecture}
\end{figure}

This section provides a brief overview of the trajectory generation and the path-following control, 
which lays the basis for  formulation of the time-coordination problem, Fig.~\ref{fig:architecture}.
\subsection{Trajectory Generation}
At the trajectory-generation level, for $n$ UAVs involved in a cooperative mission, the trajectory generation algorithm produces a set of $n$ desired collision-free trajectories
\begin{align} \label{trajectory}
    p_{d,i}(t_d): [0,t_f]\rightarrow \mathbb{R}^3, \ \ i\in\{1,\dots,n\}
\end{align}
parameterized by the mission time $t_d$. Here, $t_f$ denotes the mission duration. Considering the specifications of each UAV, the algorithm has to ensure that 
\begin{align*}
    \left\|\frac{dp_{d,i}(t_d)}{dt_d}\right\|\leq v_{d_i,max}<v_{i,max}, \\
    \left\|\frac{d^2p_{d,i}(t_d)}{dt_d^2}\right\|\leq a_{d_i,max}<a_{i,max},
\end{align*}
where $v_{i,max}$ and $a_{i,max}$ are the maximum speed and acceleration that the $i$th UAV can achieve. On the other hand, $v_{d_i,max}$ and $a_{d_i,max}$ are conservative design values. The difference between the actual dynamic limits and the conservative design values is needed to allow for variations in the pace of the mission, which will become clear in the subsequent discussion.\\ 
The mission time $t_d$ is different from the actual time $t$. In fact, the time-coordination problem can be formulated using a mapping of $t$ to $t_d$. Let $\gamma_i(t)$, referred to as the virtual time, define a map between the actual time $t$ and the mission time $t_d$ as follows:
\begin{align*}
    \gamma_i: [0,\infty)\rightarrow[0,t_f], \ \ \ i\in\{1,\dots,n\}. 
\end{align*}
Then, the position of the virtual target to be followed by the $i$th UAV is expressed as $p_{d,i}(\gamma_i(t))$. From the expression $\left\|\frac{dp_{d,i}(\gamma_i(t))}{dt}\right\|=\left\|\frac{dp_{d,i}(\gamma_i(t))}{d\gamma_i(t)} \cdot \frac{d\gamma_i(t)}{dt}\right\|=\left\|\frac{dp_{d,i}(t_d)}{dt_d} \cdot \dot{\gamma}_i(t)\right\|$, if $\dot{\gamma}_i(t)=1$, then the commanded speed $\left\|\frac{dp_{d,i}(\gamma_i(t))}{dt}\right\|$ is the \\ same as the speed profile $\left\|\frac{dp_{d,i}(t_d)}{dt_d}\right\|$ designed by the trajectory generation algorithm. On the other hand, $\dot{\gamma}_i(t)>1$ $\left(\dot{\gamma}_i(t)<1\right)$ implies a faster (slower) execution of the mission. The above discussion makes it clear that $\gamma_i(t)$ represents the progression of the $i$th UAV along its desired trajectory $p_{d,i}(\cdot)$, and $\dot{\gamma}_i(t)$ the progression rate of it. \\
The physical constraints on the speed and acceleration of each UAV lead to the following inequalities:
\begin{align}
    \left\|\frac{dp_{d,i}(\gamma_i(t))}{dt}\right\|&\leq v_{d_i,max}\dot{\gamma_i}\leq v_{i,max}, \label{c1} \\
    \left\|\frac{d^2p_{d,i}(\gamma_i(t))}{dt^2}\right\|&\leq a_{d_i,max}\dot{\gamma_i}+v_{d_i,max}\ddot{\gamma_i}\leq a_{i,max}. \label{c2}
\end{align}
Here, we can find some positive constants $\dot{\gamma}_{max}$ and $\ddot{\gamma}_{max}$ such that the following constraints
\begin{align}
    0<1-\dot{\gamma}_{max}\leq&\dot{\gamma}_i(t)\leq 1+\dot{\gamma}_{max} \label{feas1}, \\
    |&\ddot{\gamma}_i(t)|\leq\ddot{\gamma}_{max} \label{feas2}
\end{align}
imply \eqref{c1} and \eqref{c2}, respectively.

\subsection{Path Following}
In the path-following control, the path-following error is defined by $e_{PF,i}(t)\trieq p_{d,i}(\gamma_i(t))-p_i(t)$, where $p_{d,i}(\gamma_i(t))$ is the position of the $i$th virtual target and $p_i(t)$ is the actual position of the $i$th UAV. The Lyapunov-based path-following algorithm in \cite{ven2013} 
makes sure that 
\begin{align} \label{pf}
    \|e_{PF}(t)\|\leq\rho, \ \ \ \forall t\geq0,
\end{align}
where $e_{PF}(t)=[e_{PF,1}(t)^\top,\dots,e_{PF,n}(t)^\top]^\top$, and $\rho>0$ characterizes the performance of the algorithm.
\section{PROBLEM FORMULATION: TIME COORDINATION} \label{IV}
In this section, we provide rigorous descriptions of the time-coordination objectives and characterize the information flow among UAVs. Finally, we formally state the problem at hand.

As mentioned in the previous section, $\gamma_i(t)$ and $\dot{\gamma}_i(t)$ characterize the progression of the $i$th UAV along the trajectory $p_{d,i}(\cdot)$. It is said that all the UAVs involved in a cooperative mission are coordinated at time $t$, if 
\begin{align} \label{obj1}
    \gamma_i(t)=\gamma_j(t), \ \ \ \forall i,j\in\{1,\dots,n\}.
\end{align}
Furthermore, for some desired mission rate $\dot{\gamma}_d(t)>0$, if 
\begin{align} \label{obj2}
    \dot{\gamma}_i(t)=\dot{\gamma}_d(t), \ \ \ \forall i\in\{1,\dots,n\},
\end{align}
then all the UAVs are considered to be progressing with the desired mission rate. Here, $\dot{\gamma}_d(t)$ satisfies
$0<1-\dot{\gamma}_{d,max}\leq \dot{\gamma}_d(t)\leq 1+\dot{\gamma}_{d,max}$ and $|\ddot{\gamma}_d(t)|\leq\ddot{\gamma}_{d,max}$ for some constants $\dot{\gamma}_{d,max}>0$ and $\ddot{\gamma}_{d,max}>0$ to be defined in Theorem \ref{thm}.

\begin{figure} [h!]
     \centering
     \begin{subfigure}[h]{0.11\textwidth}
         \centering
         \includegraphics[width=\textwidth]{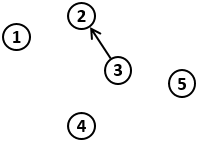}
         \caption{$\mathcal{D}_1$}
         \label{fig:g1}
     \end{subfigure}
     \hspace{0.15em}
     \begin{subfigure}[h]{0.11\textwidth}
         \centering
         \includegraphics[width=\textwidth]{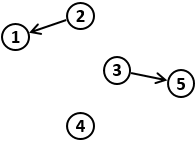}
         \caption{$\mathcal{D}_2$}
         \label{fig:g2}
     \end{subfigure}
     \hspace{0.15em}
     \begin{subfigure}[h]{0.11\textwidth}
         \centering
         \includegraphics[width=\textwidth]{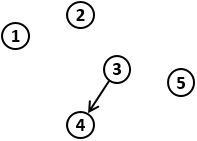}
         \caption{$\mathcal{D}_3$}
         \label{fig:g3}
     \end{subfigure}
     \hspace{0.15em}
     \begin{subfigure}[h]{0.11\textwidth}
         \centering
         \includegraphics[width=\textwidth]{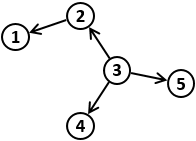}
         \caption{$\cup_{i=1}^3\mathcal{D}_i$}
         \label{fig:union}
     \end{subfigure}
        \caption{Network topologies with digraphs.}
        \label{fig:digraph}
\end{figure}

To achieve the time-coordination objectives, the UAVs are required to communicate their coordination information among themselves. The dynamic information flow is well modeled by a digraph $\mathcal{D}(t)$, whose Laplacian is denoted by $L(t)$. The following assumptions are made on the inter-UAV communication. 
\begin{assumption} \label{assum1}
The communication between two UAVs is directional with no time delays.
\end{assumption}
\begin{assumption} \label{assum2}
The $i$th UAV receives coordination information $\gamma_j(t)$ only 
from UAVs in its neighborhood set $\mathcal{N}_i(t)$.
\end{assumption}
\noindent The UAVs equipped with a transmitter can change the information flow among UAVs by changing their transmission targets. Based on this fact, we can formulate the following assumptions.
\begin{assumption} \label{assum3}
The information flow $\mathcal{D}(t)$ is switched by the UAVs equipped with a transmitter between 
$\mathcal{D}_i=\left(\mathcal{V},\mathcal{E}_i,\mathcal{A}_i\right)$, $i\in\{1,\dots,m\}$, where $\cup^{m}_{i=1}\mathcal{D}_i\trieq\left(\mathcal{V},\cup^{m}_{i=1}\mathcal{E}_i,\sum^{m}_{i=1}\mathcal{A}_i\right)$ contains a directed spanning tree. 
\end{assumption}
\noindent In this case, it is said that $\mathcal{D}_1,\dots,\mathcal{D}_m$ are jointly connected. 
An example of such $\mathcal{D}_i$'s is given in Fig. \ref{fig:digraph}. 
The Laplacian of $\cup^{m}_{i=1}\mathcal{D}_i$ is represented by $L_\cup\trieq \sum^{m}_{i=1}L_i$, where $L_i$ is the Laplacian of $\mathcal{D}_i$, $i\in\{1,\dots,m\}$. 
\begin{assumption} \label{assum4}
A switching law for $\mathcal{D}(t)$ is available to the UAVs equipped with a transmitter.
\end{assumption}
\noindent For example, in Fig. \ref{fig:digraph}, UAVs $2$, $3$ equipped with a transmitter can switch the network topology according to a switching law. 
 \\
\textit{Problem (Time-Coordination Problem):} Consider a set of $n$ UAVs assigned to the desired trajectories \eqref{trajectory}. Let the UAVs be equipped with path-following controllers that satisfy \eqref{pf}. Then, the objective is to design a decentralized coordination controller and a switching law for the inter-UAV information flow such that $\gamma_i(t)$ and $\dot{\gamma}_i(t)$ converge towards the consensus \eqref{obj1} and \eqref{obj2}, respectively, without violating the feasibility constraints \eqref{feas1} and \eqref{feas2}.

\section{MAIN RESULT} \label{V}
In this section, we introduce a coordination controller that governs the evolution of the virtual time $\gamma_i(t)$, and we design a switching law for the communication network to solve the time-coordination problem. 

Under Assumptions \ref{assum1} and \ref{assum2}, the following decentralized control law as in \cite{ven2015} can be considered
\begin{align} \label{dyn1}
    \ddot{\gamma}_i(t)&=-b(\dot{\gamma}_i(t)-\dot{\gamma}_d(t)) \nonumber \\
    &-a\sum_{j\in\mathcal{N}_i(t)}(\gamma_i(t)-\gamma_j(t))-\bar{\alpha}_i(e_{PF,i}(t)), \\
    \gamma_i(0)&=0, \ \ \dot{\gamma}_i(0)=1, \nonumber
\end{align}
where $a$ and $b$ are positive coordination control gains and $\bar{\alpha}_i(e_{PF,i}(t))$ is defined as 
\begin{align*}
    \bar{\alpha}_i(e_{PF,i}(t))=\frac{\dot{p}_{d,i}(\gamma_i(t))^\top e_{PF,i}(t)}{\|\dot{p}_{d,i}(\gamma_i(t))\|+\delta}
\end{align*}
with $\delta$ being a positive design parameter. \\
For ease of analysis, let us introduce the coordination error state $\xi_{TC}(t)=[\xi_1(t)^\top \ \xi_2(t)^\top]^\top$ with
\begin{equation}
\begin{aligned} \label{error}
    \xi_1(t)&=Q\gamma(t) \ \in \mathbb{R}^{n-1}, \\
    \xi_2(t)&=\dot{\gamma}(t)-\dot{\gamma}_d(t)1_n \ \in \mathbb{R}^{n},
\end{aligned}
\end{equation}
where $\gamma(t)=[\gamma_1(t), \dots, \gamma_n(t)]^\top$ and $Q\in\mathbb{R}^{(n-1)\times n}$ is a matrix that satisfies $Q1_n=0_{n-1}$, $QQ^\top=\mathbb{I}_{n-1}$ and $Q^\top Q=\mathbb{I}_n-\frac{1_n1_n^\top}{n}$. From the fact that the nullspace of $Q$ is spanned by $1_n$ (Lemma $7$ in \cite{phdenric2013}), if $\xi_1(t)=0_{n-1}$, then $\gamma_i(t)=\gamma_j(t)$, $\forall i,j\in\{1,\dots,n\}$. Further, $\xi_2(t)=0_n$ implies that 
$\dot{\gamma}_i(t)=\dot{\gamma}_d(t)$, $\forall i\in\{1,\dots,n\}$. Thus, $\xi_{TC}(t)=0_{2n-1}$ is equivalent to \eqref{obj1} and \eqref{obj2}. \\
The dynamics of $\gamma(t)$ is concisely rewritten as
\begin{align} 
    \ddot{\gamma}(t)&=-b\xi_2(t)-aL(t)\gamma(t)-\bar{\alpha}(e_{PF}(t)), \label{dyn2} \\
    \gamma(0)&=0_n, \ \ \dot{\gamma}(0)=1_n, \nonumber
\end{align} 
where $\bar{\alpha}(e_{PF}$ $(t))=[\bar{\alpha}_1(e_{PF,1}(t)),\dots,\bar{\alpha}_n(e_{PF,n}(t))]^\top$.
\begin{remark}
If the $i$th UAV is preceding its virtual target due to a disturbance such as tailwinds, $\bar{\alpha}_i(e_{PF,i}(t))$ in \eqref{dyn1} becomes negative, thereby accelerating $\gamma_i(t)$. It allows the UAV to fast approach the virtual target saving path-following control efforts. However, as a result, the intervehicle coordination is likely to be ruined. To resolve this situation, the second term in \eqref{dyn1} 
adjusts the evolution of $\gamma_i(t)$ in a way that the intervehicle coordination \eqref{obj1} is recovered. In the other case, where the $i$th UAV is falling behind its virtual target, the fleet restores the coordination in a similar manner. The third term in \eqref{dyn1} ensures that the fleet of UAVs progresses in accordance with the desired mission pace $\dot{\gamma}_d(t)$. 
\end{remark}

Next, we design a switching law for the inter-UAV communication network using the following lemma and a state-feedback switching law design method from \cite{zhe2005}. Under this law, the dynamics in \eqref{dyn1} can solve the time-coordination problem.
\begin{lemma} \label{lem1}
Define $\bar{L}(t)\trieq QL(t)Q^\top \in \mathbb{R}^{(n-1)\times(n-1)}$. Then, the following hold at any time $t$. \\
a) The spectrum of $\bar{L}(t)$ is the same as that of $L(t)$ without the eigenvalue $0$ whose corresponding eigenvector is $1_n$. \\
b) If $\mathcal{D}(t)$ contains a directed spanning tree, $-\bar{L}(t)$ is~Hurwitz stable. Otherwise, $-\bar{L}(t)$ is marginally stable.
\end{lemma}
\begin{proof}
a) Given $L(t)x=\lambda x$ $(x\neq 0)$, one has $QL(t)x=\lambda Qx$. 
The left hand side of the latter equation is $QL(t)x=QL(t)\left(\mathbb{I}_n-\frac{1_n1_n^\top}{n}\right)x=QL(t)Q^\top Qx=\bar{L}(t)Qx$.
Further, b) is deduced from a) and the algebraic connectivity of digraphs presented in section \ref{prelim}. 
\end{proof} 
First, we construct matrices used to design the switching law. Since $\cup^{m}_{i=1}\mathcal{D}_i$ contains a directed spanning tree (Assumption \ref{assum3}), for the Laplacian $L_\cup= \sum^{m}_{i=1}L_i$ of $\cup^{m}_{i=1}\mathcal{D}_i$,
\begin{align*}
    -\bar{L}_\cup \trieq -QL_\cup Q^\top=-\sum^{m}_{i=1}QL_iQ^\top \trieq-\sum^{m}_{i=1}\bar{L}_i
\end{align*}
is Hurwitz stable due to Lemma \ref{lem1}-b). Solving the Lyapunov equation 
\begin{align*}
    (-\bar{L}_\cup)^\top P+P(-\bar{L}_\cup)=-m\mathbb{I}_{n-1}
\end{align*}
gives a unique symmetric positive definite matrix $P$. Define
\begin{align} \label{H_i}
    H_i\trieq (-\bar{L}_i)^\top P+P(-\bar{L}_i), \ \ \ i\in\{1,\dots,m\}.
\end{align}
With $H_i$'s at hand, we formulate a state-feedback switching law for the communication network. Consider an auxiliary system whose state vector is used for designing the switching law
\begin{align} \label{aux}
    \dot{\phi}(t)=-\frac{a}{b}\bar{L}_{\sigma(t)}\phi(t), \ \ \ \phi(0)=\phi_0\neq 0,
\end{align}
where $a$ and $b$ are the coordination control gains in \eqref{dyn1}, and $\sigma(t):[0,\infty)\rightarrow\{1,\dots,m\}$ denotes the switching law to be designed, under which $L(t)$ and $\bar{L}(t)$ are equivalently rewritten as $L_{\sigma(t)}$ and $\bar{L}_{\sigma(t)}$, respectively. \\
For the given initial condition $\phi(0)=\phi_0$, the initial communication network topology is determined by
\begin{align} \label{law1}
    \sigma(t_0)=\argmin_{i\in\{1,\dots,m\}}\{\phi_0^\top H_i\phi_0\}.
\end{align}
If there are more than one such index, simply the smallest one is chosen. Now, the switching time/index sequences are recursively defined by
\begin{equation}
\begin{aligned} \label{law2}
    t_{k+1}=inf\{t>t_k : \ &\phi(t)^\top H_{\sigma(t_k)}\phi(t)> \\
    &-\mu_{\sigma(t_k)}\lambda_{max}(P)\phi(t)^\top\phi(t)\},
\end{aligned}
\end{equation}
\begin{align} \label{law3}
    \sigma(t_{k+1})=\argmin_{i\in\{1,\dots,m\}}\{\phi(t_{k+1})^\top H_i\phi(t_{k+1})\},
\end{align}
where $\mu_i\in(0,1/\lambda_{max}(P))$ and $k=0,1,2,\dots$. \\
By virtue of \eqref{law2}, it is evident that the following inequality holds
\begin{align} \label{inequ}
    \phi(t)^\top H_{\sigma(t_k)}\phi(t)\leq
    -\mu_{\sigma(t_k)}\lambda_{max}(P)\phi(t)^\top\phi(t)
\end{align}
for $t\in(t_k,t_{k+1})$, $k=0,1,2,\dots$.
\begin{lemma}
The switching law $\sigma(t)$ is well defined, i.e., the dwell time $t_{k+1}-t_k$ is lower bounded by a positive constant $\eta$ defined by
\begin{align*}
     \eta\trieq\sup_{\vartheta>1}\min_{i\in\{1,\dots,m\}}\left(\frac{1-\mu_i\lambda_{max}(P)}{\frac{a}{b}\vartheta^2\nu_i},\frac{\ln\vartheta}{\frac{a}{b}\|\bar{L}_i\|}\right),
\end{align*}
where $\nu_i\trieq\|\bar{L}_i^\top(H_i+\mathbb{I}_{n-1})+(H_i+\mathbb{I}_{n-1})\bar{L}_i\|$.
\end{lemma}
\begin{proof}       
The proof is similar to the proof of Lemma $3.26$ in \cite{zhe2005}.
\end{proof}
\begin{remark}
The switching law designed above strategically switch the topology between $\mathcal{D}_1,\dots,\mathcal{D}_m$ which are jointly connected. 
This law allows UAVs to economically use the limited communication bandwidth.
For example, as can be seen in Fig.\cref{fig:g1,fig:g2,fig:g3}, at most two communication edges are activated to coordinate five UAVs consuming a short portion of the bandwidth.


\end{remark}
\begin{remark}
As mentioned in the previous section, the UAVs can have communication devices with different specifications. Figure~\ref{fig:union} clearly illustrates this point: UAVs $1,4,5$ need a receiver; UAV $3$ needs a transmitter; UAV $2$ needs both a transmitter and a receiver. This is a clear difference from the previous work in \cite{ven2015}, where every UAV has to be equipped with both a transmitter and a receiver. 
Thus, one can cut costs on communication devices with our algorithm. Also, the information flow can be more efficiently switched as compared to the bidirectional communication case because only UAVs with a transmitter (e.g., UAVs $2,3$ in Fig. \ref{fig:digraph}) need to change their transmission targets. In the bidirectional communication case (e.g., Fig. \ref{fig:bidirectional graph}), every UAV involved in a change of topology has to change its transmission targets.
\end{remark}
The following theorem provides the main results of this paper.
\begin{theorem} \label{thm}
Consider a cooperative mission where a fleet of $n$ UAVs are assigned to the desired trajectories given in~\eqref{trajectory}. Assume that path-following controllers implemented onboard the UAVs satisfy the bound \eqref{pf}. 
Let the evolution of $\gamma_i(t)$ be governed by \eqref{dyn1} over the information flow $\mathcal{D}(t)$ switched in accordance with $\sigma(t)$ in \cref{aux,law1,law2,law3}.
Finally, let $\|\xi_{TC}(0)\|$, $\rho$, $\dot{\gamma}_{d,max}$, and $\ddot{\gamma}_{d,max}$ satisfy
\begin{align} \label{constraint1}
    \dot{\gamma}_{d,max}<\dot{\gamma}_{max}
\end{align}
and
\begin{equation}
\begin{aligned} \label{constraint2}
    \max&\{\|\xi_{TC}(0)\|,\rho,\ddot{\gamma}_{d,max}\} \\
    &\leq\min\left\{\frac{\dot{\gamma}_{max}-\dot{\gamma}_{d,max}}{\kappa_1+2\kappa_2},\frac{\ddot{\gamma}_{max}}{2b\kappa_1+4b\kappa_2+1}\right\} 
\end{aligned}
\end{equation}
for $\kappa_1$ and $\kappa_2$ defined in \eqref{kappa1} and \eqref{kappa2}, respectively. \\
Then, there exist time coordination gains 
\begin{align*}
    a>0 \ \ and \ \ b\geq\sqrt{\left(\mathcal{M}+4\mathcal{M}^2k^2_\phi/\mu+\mu/(4k^2_\phi)\right)a},
\end{align*}
where $\mathcal{M}\trieq\max_{i\in\{1,\dots,m\}}\|L_i\|$, $k_\phi\trieq\sqrt{\frac{\lambda_{max}(P)}{\lambda_{min}(P)}}$, and $\mu\trieq\min_{i\in\{1,\dots,m\}}\mu_i$ such that
\begin{equation}
\begin{aligned}
    \|\xi_{TC}(t)\|\leq\kappa_1\|\xi_{TC}&(0)\|e^{-\lambda_{TC}t} \\&+\kappa_2\sup_{t\geq0}\left(\|e_{PF}(t)\|+|\ddot{\gamma}_d(t)|\right)
\end{aligned}
\end{equation}
with rate of convergence
\begin{align} \label{rate}
    \lambda_{TC}\leq\frac{a}{6b}\frac{\mu}{k^2_\phi}.
\end{align}
Moreover, the feasibility constraints \cref{feas1,feas2} are satisfied.
\end{theorem}
\begin{proof}
To analyze the convergence properties of \eqref{dyn1}, motivated by \cite{ven2015}, we reformulate it into a stabilization problem by introducing a variable
\begin{align*}
    \chi(t)=b\xi_1(t)+Q\xi_2(t).
\end{align*}
The coordination error state $\xi_{TC}(t)=[\xi_1(t)^\top \ \xi_2(t)^\top]^\top$ can be redefined by $\bar{\xi}_{TC}(t)=[\chi(t)^\top \ \xi_2(t)^\top]^\top$ with dynamics
\begin{equation}
\begin{aligned} \label{dyn3}
    \dot{\chi}&=-\frac{a}{b}\bar{L}(t)\chi+\frac{a}{b}QL(t)\xi_2-Q\bar{\alpha}(e_{PF}) \\
    \dot{\xi}_2&=-\frac{a}{b}L(t)Q^\top\chi-\left(b\mathbb{I}_n-\frac{a}{b}L(t)\right)\xi_2-\bar{\alpha}(e_{PF})-\ddot{\gamma}_d1_n.
\end{aligned}
\end{equation}
As a step towards constructing a Lyapunov function candidate for \eqref{dyn3}, we show that the auxiliary system \eqref{aux} is globally uniformly exponentially stable (GUES). Consider $V(t)=\phi(t)^\top P \phi(t)$. Its time derivative along the trajectory of \eqref{aux} is
\begin{align*}
    \dot{V}(t)&=-\frac{a}{b}\phi^\top\left(\bar{L}^\top_{\sigma(t)}P+P\bar{L}_{\sigma(t)}\right)\phi=\frac{a}{b}\phi^\top H_{\sigma(t)}\phi \\
    &\leq-\frac{a}{b}\mu_{\sigma(t)}\lambda_{max}(P)\phi^\top\phi\leq-\frac{a}{b}\mu V(t),
\end{align*}
where the second equality is from \eqref{H_i}; the first inequality is from \eqref{inequ}, and $\sigma(t)=\sigma(t_k)$ for $t\in [t_k,t_{k+1})$; the second inequality is from  $\mu=\min_{i\in\{1,\dots,m\}}\mu_i$. Application of the comparison lemma (Lemma $3.4$ in \cite{kha2002}) yields
\begin{align*}
    V(t)\leq V(0)e^{-\frac{a}{b}\mu t}.
\end{align*} 
The system \eqref{aux} is GUES : 
\begin{align*}
    \|\phi(t)\|\leq k_\phi \|\phi(0)\|e^{-\gamma_\phi t},
\end{align*}
where $k_\phi=\sqrt{\frac{\lambda_{max}(P)}{\lambda_{min}(P)}}$ and $\gamma_\phi\trieq\frac{a}{2b}\mu$. \\
Since $\bar{L}(t)$ is continuous for almost all $t\geq0$, uniformly bounded ($\|\bar{L}(t)\|\leq\|L(t)\|\leq\max_{i\in\{1,\dots,m\}}\|L_i\|= \mathcal{M}$), and the system \eqref{aux} is GUES, a similar argument as the one in Theorem $4.12$ in \cite{kha2002} implies that for any constants $c_3$ and $c_4$ satisfying $0<c_3\leq c_4$, there exists a continuously differentiable, symmetric, positive definite matrix $\Psi(t)$ such that
\begin{align}
    c_1\mathbb{I}_{n-1}\trieq \frac{bc_3}{2a\mathcal{M}}\mathbb{I}_{n-1}\leq \Psi(t) \leq \frac{k^2_\phi c_4}{2\gamma_\phi}\mathbb{I}_{n-1}\trieq c_2\mathbb{I}_{n-1}, \label{p1} \\
    \dot{\Psi}(t)-\frac{a}{b}\bar{L}^\top(t) \Psi(t)-\frac{a}{b}\Psi(t)\bar{L}(t)\leq -c_3\mathbb{I}_{n-1}. \label{p2}
\end{align}
Now, we construct a Lyapunov function candidate for \eqref{dyn3} using $\Psi(t)$ introduced above:
\begin{align} \label{lya}
    V_{TC}(t)=\chi^\top \Psi(t)\chi+\frac{\beta}{2}\|\xi_2\|^2=\bar{\xi}^\top_{TC} W(t)\bar{\xi}_{TC},
\end{align}
where $\beta>0$ and $W(t)\trieq\begin{bmatrix}
\Psi(t) & 0 \\ 
0 & \frac{\beta}{2}\mathbb{I}_n
\end{bmatrix}$. \\
The time derivative of \eqref{lya} along the trajectory of \eqref{dyn3} is
\begin{align*}
    \dot{V}_{TC}&= \chi^\top\left(\dot{\Psi}(t)-\frac{a}{b}\bar{L}^\top(t) \Psi(t)-\frac{a}{b}\Psi(t)\bar{L}(t)\right)\chi \\
    &-\beta\xi^\top_2\left(b\mathbb{I}_n-\frac{a}{b}L(t)\right)\xi_2 \\
    &+\chi^\top\left(2\frac{a}{b}\Psi(t)QL(t)-\beta\frac{a}{b}QL^\top(t)\right)\xi_2 \\
    &-\left(2\chi^\top \Psi(t)Q+\beta\xi^\top_2\right)\bar{\alpha}(e_{PF})-\beta\xi^\top_2\ddot{\gamma}_d1_n,
\end{align*}
which leads to 
\begin{align*}
    \dot{V}_{TC}&\leq -c_3\|\chi\|^2-\beta\left(b\mathbb{I}_n-\frac{a}{b}\mathcal{M}\right)\|\xi_2\|^2 \\
    &+\left(2\frac{a}{b}\mathcal{M}\|\Psi(t)\|+\beta\frac{a}{b}\mathcal{M}\right)\|\chi\|\|\xi_2\| \\
    &+\left(2\|\Psi(t)\|\|\chi\|+\beta\|\xi_2\|\right)\left(\|\bar{\alpha}(e_{PF})\|+|\ddot{\gamma}_d|\right),
\end{align*}
where we used \eqref{p2}, $\|Q\|=1$, and $\|L(t)\|\leq \mathcal{M}$. \\
Using $\|\Psi(t)\|\leq c_2=\frac{k^2_\phi c_4}{2\gamma_\phi}=\frac{b}{a}\frac{k^2_\phi}{\mu}c_4$ in \eqref{p1} and the inequality $\|\chi\|\|\xi_2\|\leq \frac{\epsilon\|\chi\|^2}{2}+\frac{\|\xi_2\|^2}{2\epsilon}$, $\epsilon>0$, we obtain
\begin{align*}
    \dot{V}_{TC}&\leq -c_3\|\chi\|^2-\beta\left(b\mathbb{I}_n-\frac{a}{b}\mathcal{M}\right)\|\xi_2\|^2 \\
    &+\left(\frac{2\mathcal{M}k^2_\phi}{\mu}c_4+\beta\frac{a}{b}\mathcal{M}\right)\left(\frac{\epsilon\|\chi\|^2}{2}+\frac{\|\xi_2\|^2}{2\epsilon}\right) \\
    &+\left(\frac{k^2_\phi c_4}{\gamma_\phi}+\beta\right)\|\bar{\xi}_{TC}\|\left(\frac{v_{max}}{v_{min}+\delta}\|e_{PF}\|+|\ddot{\gamma}_d|\right),
\end{align*}
where $v_{max}=\max_{i}\{v_{i,max}\}$ and $v_{min}=\max_{i}\{v_{i,min}\}$.
Letting $c_3=c_4$, $\epsilon=\frac{\mu}{4\mathcal{M}k^2_\phi}$, $\beta=\frac{b}{2a\epsilon \mathcal{M}}c_4$, and $\delta>v_{max}-v_{min}$, we get the matrix form
\begin{align*}
    \dot{V}_{TC}&\leq -\bar{\xi}^\top_{TC}U\bar{\xi}_{TC}+\left(\frac{k^2_\phi c_4}{\gamma_\phi}+\beta\right)\|\bar{\xi}_{TC}\|\left(\|e_{PF}\|+|\ddot{\gamma}_d|\right),
\end{align*}
where
\begin{align*}
U\trieq\begin{bmatrix}
\frac{c_3}{2}\mathbb{I}_{n-1} & 0 \\ 
0 & \beta\left(b-\frac{a}{b}\mathcal{M}-\frac{4\mathcal{M}^2k^2_\phi}{\mu}\frac{a}{b}\right)\mathbb{I}_n
\end{bmatrix}.
\end{align*}
We let $b\geq\sqrt{\left(\mathcal{M}+\frac{4\mathcal{M}^2k^2_\phi}{\mu}+\frac{\mu}{4k^2_\phi}\right)a}$ and $\lambda_{TC}\leq\frac{c_3}{6c_2}=\frac{a}{6b}\frac{\mu}{k^2_\phi}$ so that the following inequality holds:
\begin{align*}
    &U-3\lambda_{TC}W \\
    &\resizebox{\hsize}{!}{$\geq
    \begin{bmatrix}
    \left(\frac{c_3}{2}-3\lambda_{TC}c_2\right)\mathbb{I}_{n-1} & 0 \\ 
    0 & \beta\left(b-\frac{a}{b}\mathcal{M}-\frac{4\mathcal{M}^2k^2_\phi}{\mu}\frac{a}{b}-\frac{3}{2}\lambda_{TC}\right)\mathbb{I}_n
    \end{bmatrix}\geq 0.
    $}
\end{align*}
The derivative of $V_{TC}$ is upper bounded by
\begin{align*}
    \dot{V}_{TC}&\leq -3\lambda_{TC}V_{TC}+\left(\frac{k^2_\phi c_4}{\gamma_\phi}+\beta\right)\|\bar{\xi}_{TC}\|\left(\|e_{PF}\|+|\ddot{\gamma}_d|\right) \\
    &\leq-2\lambda_{TC}V_{TC}-\lambda_{TC}\min\{c_1,\beta/2\}\|\bar{\xi}_{TC}\|^2 \\
    &+\left(\frac{k^2_\phi c_4}{\gamma_\phi}+\beta\right)\|\bar{\xi}_{TC}\|\left(\|e_{PF}\|+|\ddot{\gamma}_d|\right).
\end{align*}
Applying Lemma $4.6$ in \cite{kha2002} and the state transformation
$\bar{\xi}_{TC}=S\xi_{TC}\trieq
\begin{bmatrix}
b\mathbb{I}_{n-1} & Q \\ 
0 & \mathbb{I}_n
\end{bmatrix}\xi_{TC}$,
we can conclude that
\begin{equation} \label{iss}
\begin{aligned}
    \|\xi_{TC}(t)\|\leq\kappa_1\|\xi_{TC}&(0)\|e^{-\lambda_{TC}t} \\&+\kappa_2\sup_{t\geq0}\left(\|e_{PF}(t)\|+|\ddot{\gamma}_d(t)|\right),
\end{aligned}
\end{equation}
where
\begin{align}
    \kappa_1&\trieq\|S^{-1}\|\sqrt{\frac{\max\{c_2,\beta/2\}}{\min\{c1,\beta/2\}}}\|S\|, \label{kappa1} \\
    \kappa_2&\trieq\|S^{-1}\|\sqrt{\frac{\max\{c_2,\beta/2\}}{\min\{c1,\beta/2\}}}\frac{\frac{k^2_\phi c_4}{\gamma_\phi}+\beta}{\lambda_{TC}\min\{c_1,\beta/2\}}. \label{kappa2} 
\end{align}
Lastly, it can be shown that $\dot{\gamma}_i(t)$ and $\ddot{\gamma}_i(t)$ $\forall i\in\{1,\dots,n\}$ satisfy the feasibility constraints \cref{feas1,feas2} from the assumptions \eqref{constraint1} and \eqref{constraint2}. 
From the inequality $|\dot{\gamma}_i(t)-1|\leq|\dot{\gamma}_d(t)-1|+|\dot{\gamma}_i(t)-\dot{\gamma}_d(t)|$ and \eqref{iss}, it follows that
\begin{align*}
    |\dot{\gamma}_i(t)-1|\leq|\dot{\gamma}_d(t)-1|+\kappa_1&\|\xi_{TC}(0)\|e^{-\lambda_{TC}t} \\
    &+\kappa_2\sup_{t\geq0}\left(\|e_{PF}(t)\|+|\ddot{\gamma}_d(t)|\right).
\end{align*}
Recalling $|\dot{\gamma}_d(t)-1|<\dot{\gamma}_{d,max}$, $|\ddot{\gamma}_d(t)|\leq\ddot{\gamma}_{d,max}$ and \eqref{pf}, the inequality is written as
\begin{align*}
    |\dot{\gamma}_i(t)-1|&\leq\dot{\gamma}_{d,max}+\kappa_1\|\xi_{TC}(0)\|+\kappa_2\rho+\kappa_2\ddot{\gamma}_{d.max} \\
    &\leq\dot{\gamma}_{d,max}+\left(\kappa_1+2\kappa_2\right)\max\{\|\xi_{TC}(0)\|,\rho,\ddot{\gamma}_{d.max}\}.
\end{align*}
Finally, the assumptions \eqref{constraint1} and \eqref{constraint2} lead us to the conclusion that \eqref{feas1} holds. Now we consider bounds on $\ddot{\gamma}_i(t)$. From \eqref{dyn2}, it is shown that
\begin{align*}
    |\ddot{\gamma}_i(t)|&\leq b\|\xi_2(t)\|+a\mathcal{M}\|\xi_1(t)\|+\|e_{PF}(t)\| \\
    &\leq 2b\|\xi_{TC}(t)\|+\|e_{PF}(t)\|,
\end{align*}
where the second inequality is obtained by setting $b\geq a\mathcal{M}$. Recalling \eqref{iss}, it is seen that $\ddot{\gamma}_i(t)$ is bounded by
\begin{align*}
    |\ddot{\gamma}_i(t)|\leq \left(2b\kappa_1+4b\kappa_2+1\right)\max\{\|\xi_{TC}(0)\|,\rho,\ddot{\gamma}_{d.max}\}.
\end{align*}
The above inequality, together with \eqref{constraint1} and \eqref{constraint2}, implies that \eqref{feas2} holds, which completes the proof of Theorem 1.
\end{proof}
\begin{remark}
Notice that $\mu_i$, $i\in\{1,\dots,m\}$ are tunable parameters. Large values of those decrease the switching threshold $-\mu_{\sigma(t_k)}\lambda_{max}(P)\phi(t)^\top\phi(t)$ in \eqref{law2}, thereby 
increasing switching frequency. Also, the rate of convergence \eqref{rate} is increased because it is proportional to $\mu=\min_{i\in\{1,\dots,m\}}\mu_i$. 
\end{remark}
\section{SIMULATION RESULTS} \label{VI}
\begin{figure} [h!]
    \centering
    \includegraphics[width = 1.00\linewidth]{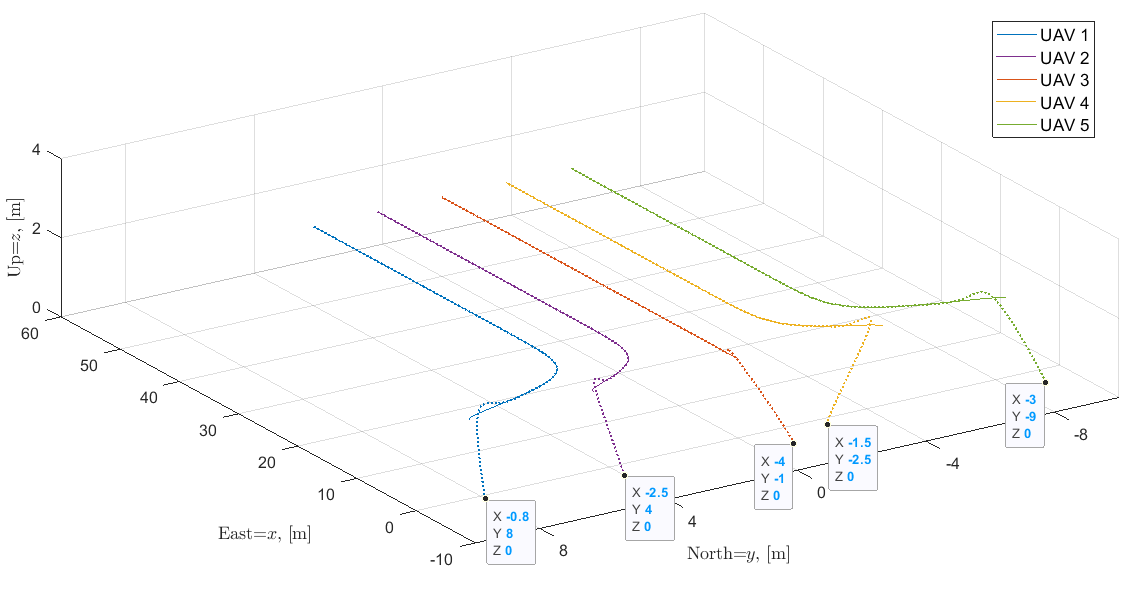}
    \caption{Time-coordinated path following of five UAVs.}
    \label{fig:traj}
\end{figure}

\begin{figure} [h!]
    \centering
    \includegraphics[width = \linewidth]{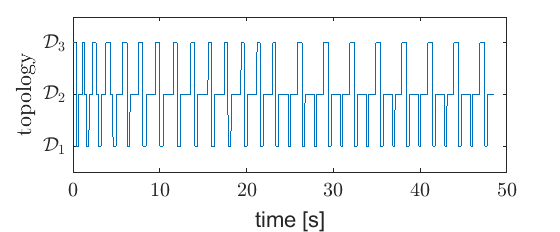}
    \caption{Evolution of $\mathcal{D}(t)$.}
    \label{fig:topology}
\end{figure}

\begin{figure} [h!]
    \centering
    \includegraphics[width = \linewidth]{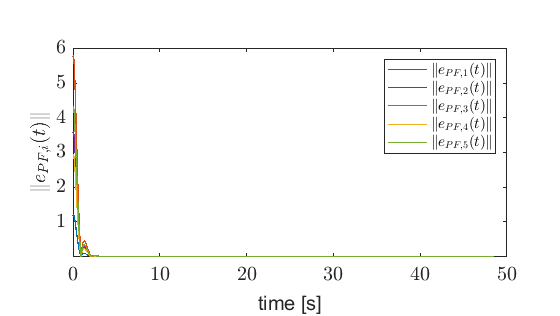}
    \caption{Convergence of the path-following errors to $0$.}
    \label{fig:PF_error}
\end{figure}

\begin{figure} [h!]
    \centering
    \includegraphics[width = \linewidth]{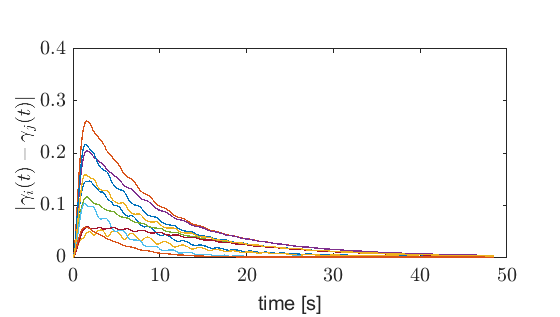}
    \caption{Convergence of the errors between virtual times to 0.}
    \label{fig:gamma}
\end{figure}

\begin{figure} [h!]
    \centering
    \includegraphics[width = \linewidth]{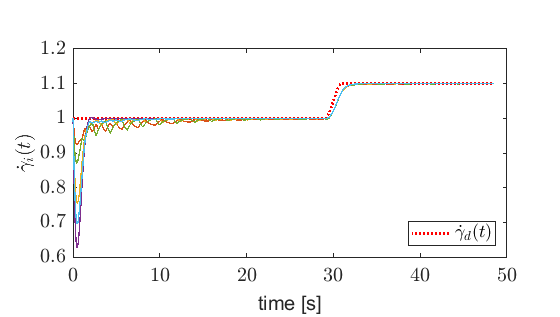}
    \caption{Evolution of the mission pace which tracks the desired mission pace $\dot{\gamma}_d(t)$.}
    \label{fig:gamma_dot}
\end{figure}

This section demonstrates that the time coordination of multiple UAVs can be achieved by the coordination control law \eqref{dyn1} over the directional inter-UAV information flow $\mathcal{D}(t)$ switched in accordance with $\sigma(t)$ in \cref{aux,law1,law2,law3}. We~also show that the proposed algorithm achieves it with significantly reduced inter-UAV communication as compared to the previous work \cite{ven2015}. \\
Let us consider a coordinated path-following mission where five UAVs are involved. The dynamics and the path-following controller implemented onboard are given in \cite{ven2013}.
The desired trajectories assigned to them are
\begin{align} \label{traj}
    p_{d,i}(t_d):[0,50]\rightarrow\begin{bmatrix}
t_d\\ 
d_i-e^{-0.6t_d}(5+3t_d)\sin{\theta_i}\\ 
2
\end{bmatrix}, 
\end{align}
where $d_i=6-2i$ and $\theta_i=-\pi/2+\pi i/6$ for $i\in\{1,\dots,5\}$. Figure \ref{fig:traj} depicts the desired trajectories (solid lines) and the paths tracked by the UAVs (dotted lines). Initially, the UAVs are on the ground and discoordinated. The control gains and tunable parameters are chosen as $a=0.75$, $b=1.82$, $\delta=1.2$, $\phi_0=[0.9, 1.7, 1.1, 0.1]^\top$, and $\mu_1=\mu_2=\mu_3=0.2638$.
 \\
In this mission, the UAVs are tasked to reconnoiter the area $-4\leq y\leq4$ keeping abreast of one another and arrive at $x=50$ simultaneously. To be more specific, with the initial path-following errors, they have to quickly catch up with their virtual targets, achieve intervehicle coordination and maintain it until the end of the mission. Additionally, they have to progress in accordance with the desired mission~pace $\dot{\gamma}_d(t)$ depicted as the dotted line in Fig. \ref{fig:gamma_dot}. \\ 
At any time $t$, the communication network $\mathcal{D}(t)$ is character-\\ized by one of $\mathcal{D}_1$, $\mathcal{D}_2$ and $\mathcal{D}_3$ depicted in Fig. \cref{fig:g1,fig:g2,fig:g3}. Notice that none of $\mathcal{D}_1$, $\mathcal{D}_2$ and $\mathcal{D}_3$ contains a directed spanning tree. Only $\cup_{i=1}^3\mathcal{D}_i$ in Fig. \ref{fig:union} is required to contain a directed spanning tree in our algorithm. Figure~\ref{fig:topology} shows the evolution of the network topology under the switching law $\sigma(t)$ given in \cref{aux,law1,law2,law3} as the mission unfolds. Considering that $\mathcal{D}_1$, $\mathcal{D}_2$ and $\mathcal{D}_3$ are not connected, the network is not connected at all times throughout the mission.
 \\
It is illustrated in Fig. \ref{fig:gamma} and Fig. \ref{fig:gamma_dot}  how the coordination dynamics \eqref{dyn1} works to solve the time-coordination problem. Since the desired trajectories start at $x=0$ and the UAVs lie in the left hand side of $x=0$ at $t=0$, they are initially put behind the schedule. This causes $\bar{\alpha}_i(x_{PF,i}(0))$ in \eqref{dyn1} to be positive, which leads to deceleration of $\gamma_i(t)$ right after the mission unfolds, as seen in Fig. \ref{fig:gamma_dot}. By virtue of it, the UAVs are allowed to fast approach their virtual targets saving path-following control efforts, Fig. \ref{fig:PF_error}. However, different sizes of deceleration destroy the coordination $\gamma_i(t)=\gamma_j(t)$. To fix it, the second term in \eqref{dyn1} adjusts the evolution of $\gamma_i(t)$ in a way that, as shown in Fig. \ref{fig:gamma}, $|\gamma_i(t)-\gamma_j(t)|$ converges to $0$. When some or all of the UAVs are deviated from their virtual targets in the middle of the mission by wind gusts, they can recover the coordination in the same manner. The effect of the first term in \eqref{dyn1} allows the UAVs to progress in accordance with the desired mission pace $\dot{\gamma}_d(t)$. In Fig. \ref{fig:gamma_dot}, it is shown that the UAVs quickly adjust their pace to match the increase in $\dot{\gamma}_d(t)$ at $t\approx30$s. Even though there was a decrease in the mission pace due to the initial path-following errors, the increase in the desired mission pace $\dot{\gamma}_d(t)$ at $t\approx30$s gets the UAVs to arrive at their final destination $x=50$ at $t=48.55$s, which is a little bit earlier than the original schedule \eqref{traj}.

\begin{figure} [h!]
     \centering
     \begin{subfigure}[h]{0.11\textwidth}
         \centering
         \includegraphics[width=\textwidth]{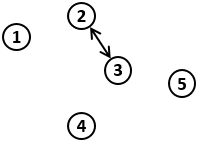}
         \caption{$\mathcal{G}_1$}
         \label{fig:bg1}
     \end{subfigure}
     \hspace{0.15em}
     \begin{subfigure}[h]{0.11\textwidth}
         \centering
         \includegraphics[width=\textwidth]{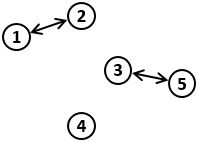}
         \caption{$\mathcal{G}_2$}
         \label{fig:bg2}
     \end{subfigure}
     \hspace{0.15em}
     \begin{subfigure}[h]{0.11\textwidth}
         \centering
         \includegraphics[width=\textwidth]{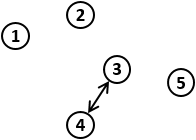}
         \caption{$\mathcal{G}_3$}
         \label{fig:bg3}
     \end{subfigure}
     \hspace{0.15em}
     \begin{subfigure}[h]{0.11\textwidth}
         \centering
         \includegraphics[width=\textwidth]{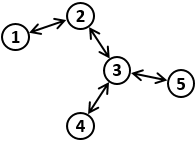}
         \caption{$\cup_{i=1}^3\mathcal{G}_i$}
         \label{fig:bg_union}
     \end{subfigure}
        \caption{Network topologies with bidirectional graphs.}
        \label{fig:bidirectional graph}
\end{figure}

Lastly, it is demonstrated that our algorithm can solve the time-coordination problem with substantially reduced inter-UAV communication as compared to the previous work \cite{ven2015}. Let us reconsider the above coordinated path-following scenario where everything is the same except the communication network is now a bidirectional graph $\mathcal{G}(t)$. The topology at time $t$ is characterized by one of Fig. \cref{fig:bg1,fig:bg2,fig:bg3}.     

\begin{figure} [h!]
    \centering
    \includegraphics[width = \linewidth]{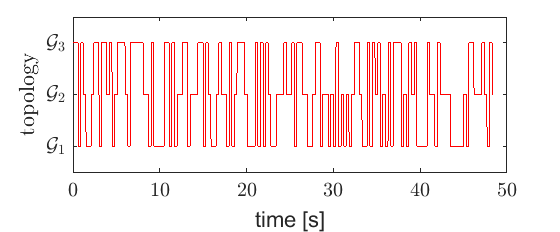}
    \caption{Evolution of $\mathcal{G}(t)$.}
    \label{fig:topology_bg}
\end{figure}

\begin{figure} [h!]
    \centering
    \includegraphics[width = \linewidth]{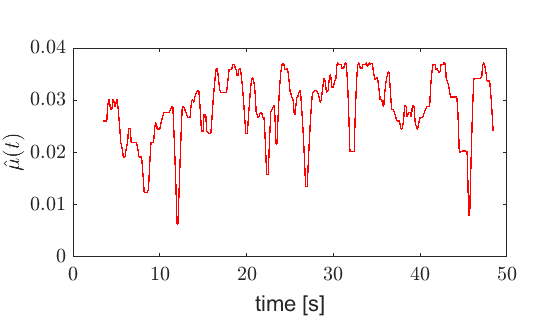}
    \caption{Connectedness of $\mathcal{G}(t)$ in an integral sense.}
    \label{fig:qos}
\end{figure}

\begin{figure} [h!]
    \centering
    \includegraphics[width = \linewidth]{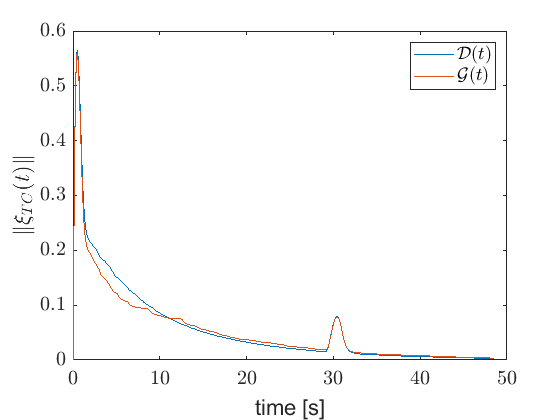}
    \caption{Time-coordination performances over the directed network $\mathcal{D}(t)$ and the bidirectional network $\mathcal{G}(t)$.}
    \label{fig:error}
\end{figure}

\noindent The network topology is randomly switched every $0.3$s as in Fig. \ref{fig:topology_bg}. It is evident from Fig. \ref{fig:qos} that  
\begin{align*}
    \hat{\lambda}(t)\trieq\lambda_{min}\left(\frac{1}{nT}\int_{t-T}^{t}QL(\tau)Q^\top d\tau\right)\geq \hat{\lambda}_{min}>0, \ t\geq T,
\end{align*}
with $n=5$, $T=3.4$s and $\hat{\lambda}_{min}=0.0062$.
It means that $\mathcal{G}(t)$ is connected in an integral sense even though it is not connected pointwise in time during the mission. 
This PE-like condition was presented in the previous work \cite{ven2015} as a sufficient condition on the network connectivity for achieving the time-coordination objectives.
 \\
Figure \ref{fig:error} shows the evolution of the norm of the coordination error state $\xi_{TC}(t)=[\xi_1(t)^\top \ \xi_2(t)^\top]^\top$ in \eqref{error} when the coordinated path-following mission unfolds over $\mathcal{D}(t)$ switched as in Fig. \ref{fig:topology} and over $\mathcal{G}(t)$ switched as in Fig. \ref{fig:topology_bg}.   
 \\
With the similar time-coordination performances observed in Fig. \ref{fig:error}, let us compare the amount of required inter-UAV communication during the mission. 
As the Adjacency matrix $\mathcal{A}(t)$ characterizes the information flow among the UAVs at time $t$, $\sum_{i,j=1}^{n}\left[\int_{0}^{\tau_f}\mathcal{A}(t)dt\right]_{ij}$ quantifies the entire amount of inter-UAV information flow during the mission. Here, $\tau_f$ denotes the time instant when the UAVs arrive at their final destinations, the values of $\tau_f$ over $\mathcal{D}(t)$ and $\mathcal{G}(t)$ are $48.55$s and $48.42$s, respectively.
\begin{table}[h!] 
\begin{center}  
\begin{tabular}{|c||c||c|} 
\hline
Amount of inter-UAV communication & $\mathcal{D}(t)$ & $\mathcal{G}(t)$ \\
\hline
$\sum_{i,j=1}^{n}\left[\int_{0}^{\tau_f}\mathcal{A}(t)dt\right]_{ij}$ & 77.17 & 126.48 \\
\hline
\end{tabular}
\end{center}
\caption{Comparison of the amount of required inter-UAV communication.} \label{table_example}
\end{table} \\
It is demonstrated in TABLE \ref{table_example} that our time-coordination algorithm solves the problem with less amount of inter-UAV communication as compared to the previous algorithm \cite{ven2015}.  

\section{CONCLUSION} \label{VII}
In this paper, a time-coordination algorithm is developed for multi-UAV cooperative missions where the communication between UAVs is not required to be bidirectional. We design a switching law for the inter-UAV information flow, over which it is shown that a decentralized coordination controller achieves the time coordination objectives. Finally, the simulation results demonstrate the efficacy of the algorithm.



\bibliographystyle{ieeetr}
\bibliography{references}

\begin{thebibliography}{10}

\bibitem{hye2017}
H.~Lee and H.~J. Kim, ``Constraint-based cooperative control of multiple aerial
  manipulators for handling an unknown payload,'' {\em IEEE Transactions on
  Industrial Informatics}, vol.~13, no.~6, pp.~2780--2790, 2017.

\bibitem{tae2018}
T.~Lee, ``Geometric control of quadrotor {UAV}s transporting a cable-suspended
  rigid body,'' {\em IEEE Transactions on Control Systems Technology}, vol.~26,
  no.~1, pp.~255--264, 2018.

\bibitem{isa2016}
I.~Deutsch, M.~Liu, and R.~Siegwart, ``A framework for multi-robot pose graph
  {SLAM},'' in {\em IEEE International Conference on Real-time Computing and
  Robotics}, pp.~567--572, 2016.

\bibitem{pat2017}
P.~Schmuck and M.~Chli, ``Multi-{UAV} collaborative monocular {SLAM},'' in {\em
  IEEE International Conference on Robotics and Automation}, pp.~3863--3870,
  2017.

\bibitem{tri2018}
T.~Sherman, J.~Tellez, T.~Cady, J.~Herrera, H.~Haideri, J.~Lopez, M.~Caudle,
  S.~Bhandari, and D.~Tang, ``Cooperative search and rescue using autonomous
  unmanned aerial vehicles,'' in {\em AIAA Information Systems-AIAA Infotech
  @Aerospace}, 2018.

\bibitem{sat2017}
S.~G. Manyam, S.~Rasmussen, D.~W. Casbeer, K.~Kalyanam, and S.~Manickam,
  ``Multi-{UAV} routing for persistent intelligence surveillance \&
  reconnaissance missions,'' in {\em International Conference on Unmanned
  Aircraft Systems}, pp.~573--580, 2017.

\bibitem{cho2016}
R.~Choe, J.~Puig-Navarro, V.~Cichella, E.~Xargay, and N.~Hovakimyan,
  ``Cooperative trajectory generation using {P}ythagorean hodograph
  {B}\'{e}zier curves,'' {\em Journal of Guidance, Control, and Dynamics},
  vol.~39, no.~8, pp.~1744--1763, 2016.

\bibitem{aug2012}
F.~Augugliaro, A.~P. Schoellig, and R.~D'Andrea, ``Generation of collision-free
  trajectories for a quadcopter fleet: A sequential convex programming
  approach,'' in {\em International Conference on Intelligent Robots and
  Systems}, pp.~1917--1922, 2012.

\bibitem{lap2006}
L.~Lapierre, D.~Soetanto, and A.~Pascoal, ``Non-singular path-following control
  of a unicycle in the presence of parametric modeling uncertainties,'' {\em
  International Journal of Robust and Nonlinear Control}, vol.~16, no.~10,
  pp.~485--505, 2006.

\bibitem{gha2007}
R.~Ghabcheloo, {\em Coordinated Path Following of Multiple Autonomous
  Vehicles}.
\newblock PhD thesis, Technical University of Lisbon, 2007.

\bibitem{ven2011}
V.~Cichella, I.~Kaminer, V.~Dobrokhodov, E.~Xargay, N.~Hovakimyan, and
  A.~Pascoal, ``Geometric 3{D} path-following control for a fixed-wing {UAV} on
  {SO}(3),'' in {\em AIAA Guidance, Navigation, and Control Conference}, 2011.

\bibitem{xar2013}
E.~Xargay, I.~Kaminer, A.~Pascoal, N.~Hovakimyan, V.~Dobrokhodov, V.~Cichella,
  A.~P. Aguiar, and R.~Ghabcheloo, ``Time-critical cooperative path following
  of multiple unmanned aerial vehicles over time-varying networks,'' {\em
  Journal of Guidance, Control, and Dynamics}, vol.~36, no.~2, pp.~499--516,
  2013.

\bibitem{ven2015}
V.~Cichella, I.~Kaminer, V.~Dobrokhodov, E.~Xargay, R.~Choe, and N.~Hovakimyan,
  ``Cooperative path following of multiple multirotors over time-varying
  networks,'' {\em IEEE Transactions on Automation Science and Engineering},
  vol.~12, no.~3, pp.~945--957, 2015.

\bibitem{pui2015}
J.~Puig-Navarro, E.~Xargay, R.~Choe, and N.~Hovakimyan, ``Time-critical
  coordination of multiple {UAV}s with absolute temporal constraints,'' in {\em
  AIAA Guidance, Navigation, and Control Conference}, 2015.

\bibitem{cam2020}
C.~Tabasso, V.~Cichella, S.~B. Mehdi, T.~Marinho, and N.~Hovakimyan,
  ``Guaranteed collision avoidance in multivehicle cooperative missions using
  speed adjustment,'' {\em Journal of Aerospace Information Systems}, vol.~17,
  no.~8, pp.~436--453, 2020.

\bibitem{cam2021}
C.~Tabasso, V.~Cichella, S.~B. Mehdi, T.~Marinho, and N.~Hovakimyan, ``Time
  coordination and collision avoidance using leader-follower strategies in
  multi-vehicle missions,'' {\em Robotics}, vol.~10, no.~1, p.~34, 2021.

\bibitem{gua2005}
G.~Xie and L.~Wang, ``Consensus control for networks of dynamic agents via
  active switching topology,'' in {\em International Conference on Natural
  Computation}, pp.~424--433, 2005.

\bibitem{yan2018a}
Y.~Mao and Z.~Zhang, ``Second-order consensus for multi-agent systems by
  state-dependent topology switching,'' in {\em American Control Conference},
  pp.~3392--3397, 2018.

\bibitem{yan2018b}
Y.~Mao, E.~Akyol, and Z.~Zhang, ``Second-order consensus for multi-agent
  systems by time-dependent topology switching,'' in {\em IEEE Conference on
  Decision and Control}, pp.~6151--6156, 2018.

\bibitem{ert2017}
E.~N. Ciftcioglu, S.~Pal, K.~S. Chan, D.~H. Cansever, A.~Swami, A.~K. Singh,
  and P.~Basu, ``Topology design games and dynamics in adversarial
  environments,'' {\em IEEE Journal on Selected Areas in Communications},
  vol.~35, no.~3, pp.~628--642, 2017.

\bibitem{maz2011}
S.~K. Mazumder, {\em Wireless networking based control}.
\newblock Springer, 2011.

\bibitem{zhe2005}
Z.~Sun, {\em Switched Linear Systems: Control and Design}.
\newblock Springer-Verlag London, 2005.

\bibitem{ven2013}
V.~Cichella, R.~Choe, S.~B. Mehdi, E.~Xargay, N.~Hovakimyan, I.~Kaminer, and
  V.~Dobrokhodov, ``A 3{D} path-following approach for a multirotor {UAV} on
  {SO}(3),'' {\em IFAC Proceedings Volumes}, vol.~46, no.~30, pp.~13--18, 2013.

\bibitem{phdenric2013}
E.~Xargay, {\em Time-Critical Cooperative Path-Following Control of Multiple
  Unmanned Aerial Vehicles}.
\newblock PhD thesis, University of Illinois at Urbana-Champaign, 2013.

\bibitem{kha2002}
H.~K. Khalil, {\em Nonlinear Systems}.
\newblock Prentice-Hall, Englewood Cliffs, NJ, 2002.

\end{thebibliography}

\end{document}